\documentclass[showpacs,twocolumn,amsmath,amssymb,pre,superscriptaddress,letterpaper]{revtex4}\newcommand{\switchsize}[0]{0.66}
\usepackage{epsfig}
\usepackage{graphicx}
\usepackage{dcolumn}

\begin{document}
\title{Dynamical heterogeneity in a glass forming ideal gas}
\author{Patrick Charbonneau}
    \affiliation{FOM Institute for Atomic and Molecular Physics, Kruislaan 407, 1098 SJ Amsterdam, The
    Netherlands}
\author{Chinmay Das}
    \affiliation{Department of Physics and Astronomy, University of Leeds, Leeds LS2 9JT, UK}
\author{Daan Frenkel}
    \affiliation{FOM Institute for Atomic and Molecular Physics, Kruislaan 407, 1098 SJ Amsterdam, The Netherlands}
\date{\today}

\begin{abstract}
We conduct a numerical study of the dynamical behavior of a system of three-dimensional ``crosses'', particles that consist of three mutually perpendicular line segments of length $\sigma$ rigidly joined at their midpoints. In an earlier study [W.~van~Ketel~{\em et~al.}, Phys. Rev. Lett. {\bf 94}, 135703 (2005)] we showed that this model has the structural properties of an ideal gas, yet the dynamical properties of a strong glass former.  In the present paper we report an extensive study of the dynamical heterogeneities that appear in this system in the regime where glassy behavior sets in.  On the one hand, we find that the propensity of a particle to diffuse is determined by the structure of its local environment. The local density around mobile particles is significantly less than the average density, but there is little clustering of mobile particles, and the clusters observed tend to be small. On the other hand, dynamical susceptibility results indicate that a large dynamical length scale develops even at moderate densities. This suggests that propensity and other mobility measures are an incomplete measure of dynamical length scales in this system. \end{abstract}
\maketitle

\section{Introduction}
There exist a bewildering variety of theories for the glass transition (see e.g.~\cite{adams-gibbs65,cohen:59,EdwardsVilgis,gotze:91,gotze:91b,
schilling:rev:03,reichman:05,Kivelson,butler:1991,butler:1991b,Chandler,Berthier,zaccarelli01}). Roughly speaking, one can distinguish between two main classes. Theories belonging to the first class are based on the assumption that static, structural correlations in the fluid are ultimately responsible for the occurrence of structural arrest. Theories that belong to the second class assume that purely kinetic factors control the onset of glassy behavior. It is probably fruitless to search for the ``true'' theory of the glass transition, because not all experimental glasses appear to be equivalent~\cite{ghosh96,Catesetal}. However, it is important to disentangle, as much as possible, the roles of structural correlations and of purely kinetic effects in the absence of such correlations.

Recently, we reported simulations that provided evidence that it is possible to observe glassy behavior in a  model system that has the structural properties of an ideal gas~\cite{ketel:05}. As the particles in an ideal gas have no static structural correlations, dynamical arrest in this system is a purely kinetic effect. The model system we explore consists of particles made of three mutually perpendicular line segments of length $\sigma$, rigidly joined at their midpoints. These three-dimensional ``crosses'' generalize the hard-needle model developed to study topological effects on rotational and translational diffusion~\cite{frenkel81,frenkel81b}, as has already been implicitly~\cite{EdwardsVilgis} or explicitly~\cite{Obukhov} suggested. A lattice-based version of the hard-needle system has already been studied by several groups as a model for orientational glass formers~\cite{renner:95,jimenez:02,schilling:rod,schilling:rodb}. Renner {\em et~al}.~\cite{renner:95} simulated line segments that can rotate around fixed lattice points. The system enters a non-ergodic glassy phase at finite segment length, but since it has an ideal static behavior the standard mode-coupling theory (MCT) of the glass transition is inapplicable. However, an extension of MCT that includes torque-torque contributions does predict a glass transition for these lattice rotators~\cite{schilling:rod,schilling:rodb}. Closer to the present model is the thin line segments with fixed but random orientations, whose dynamics was studied by Szamel {\em et~al}.~\cite{szamel:93,szamel:93b}. Using a mean-field approximation, they found that the transverse motion of the line segments decreases severely with increasing segment length, because of entanglement (tube constraints), while the motion along the
orientation of the lines is not affected by such constraint.

Since the crosses have zero volume and thus zero excluded volume, {\em all} static thermodynamic quantities are exactly known. By random insertion one can trivially generate a representative equilibrium configuration at any density. As our model is an ideal gas, the onset of glassy behavior takes place within a single thermodynamically stable phase. Hence we need not worry that the dynamics in the glassy phase be obscured by the slow nucleation of another phase. We can also safely ignore the ``Kauzmann paradox''~\cite{Kauzmann}, which states that the glass transition takes place when the entropy of the fluid phase threatens to drop below that of the crystal phase. The present model has no crystal phase nor for that matter any ordered phase. Nonetheless, its dynamics is highly nontrivial.

The rest of the paper is organized as follows. In Sec.~\ref{sec:sim_algo} we describe the model and the simulation algorithm. Collision as well as diffusion properties are presented in Sec.~\ref{sec:res:bulk} and the self-intermediate scattering function in Sec.~\ref{sec:fsqt}. We investigate in details the effect of the local environment on the mobility of particles and clustering of the ``mobile'' particles in Sec.~\ref{sec:dyn_het}, while Sec.~\ref{sec:chi_4} concerns itself with the density and wave-vector dependence of the four-point susceptibility. Finally, we conclude in Sec.~\ref{sec:conclu} with a summary of the important findings.

\section{Simulation technique \label{sec:sim_algo}}

We simulate a three-dimensional system at constant number of particles $N$, volume $V$ and temperature $T$ under Newtonian dynamics. The particles consist of three mutually perpendicular line segments of length $\sigma$ rigidly joined at their midpoints. We choose the initial center of mass positions of the particles at random in the cubic simulation box. The box volume $V = N \sigma^3/\rho$ is set by the choice of $N$ and the number density $\rho$. The cross orientations are also randomly distributed. For numerical convenience we reject configurations having two crosses with almost identical orientations to within an angle of $10^{-4}$ radians. Assuming truly random orientations, the probability  of having such closely aligned pairs of crosses is less than one part in $10^{5}$ for the system sizes considered. Hence, the effect of this choice should be negligible. We also neglect rotational motion, which correspond to having crosses with an infinite moment of inertia, so they preserve their initial orientation throughout the simulation. This allows us to analytically compute the time before the next collision, which leads to large computational efficiency gains. The initial velocities are randomly drawn from a Maxwell-Boltzmann distribution and shifted to set the center of mass velocity to zero. We choose $\sigma$ as the unit of length, the thermal energy $k_B T$ as the unit of energy, and the particle mass $m$ as the unit of mass. This results in time $t$ being expressed in units of $(k_BT/m\sigma^2)^{-1/2}$. Simple periodic boundary conditions are used in all three directions. The dynamical rules are simple: between collisions, the particles move ballistically, while when two line segments collide, the component of relative velocity perpendicular to the plane of the two line segments is reversed.

We use an event-driven algorithm~\cite{rapaport}, wherein future collision events are stored in a binary tree structure and the particle positions are updated asynchronously to the time of the next collision event. Because of the extreme anisotropy of the crosses, both spherical neighbor lists and cell structures are inefficient at high densities. Instead we consider spherocylinders around each line segment and create neighbor lists from the spherocylinder overlaps. ``Events'' in our algorithm are not only collisions, but also neighbor list updates. These occur whenever  the center of mass of a particular cross has moved by more than half the spherocylinder radius since that cross's neighbor list was last created. To limit the search for spherocylinder overlaps while creating the neighbor lists, we consider a cubic cell structure based on the center of mass of the crosses. We limit the size of the event tree by setting a time (typically five times what it would take a particle with the average speed to ballistically cross the neighbor list cutoff length) beyond which events are not entered in the tree structure, which also sets the longest survival time of a neighbor list. This ensures that if a particular cross does not undergo any collision within this interval we still correctly identify future events that involves it.

For the very rare case of quasi-simultaneous collisions, the behavior of the program is unpredictable. Depending on the exact sequence of instructions, a future event can behave like a past event and vice versa. We avoid this problem by discarding events that are separated by less than $10^{-14}$ time units from a previous event. Since this time is much smaller than the average time between  collisions even at the highest density considered in this study, this artificial exclusion does not affect the statistical analysis of our data.

After a collision, all events involving the colliding pairs are removed from the event tree and new future events are generated from their respective neighbor lists. When the event is a neighbor list update, all events involving this particle are removed and the list is recreated anew. When the next event is later than the time at which we are required to calculate any property of the system, we update the positions and velocities of all the particles to  that time without changing the event list, since by definition the next event is later than this time.

For the highest densities and largest system sizes considered
in this work, the number of collision events  in a single run
often exceeds $10^{10}$. On a 2~GHz AMD Opteron linux desktop
using an Intel Fortran compiler, the CPU time required for
$10^{10}$ collisions to take place in a system of $4096$
crosses at $\rho=20$ is about 25 hours.  The two most costly
operations are finding the future collisions and filling the
spherocylindrical neighbor list. The optimum performance is
observed when the (density dependent) spherocylinder radius is
chosen such that the average number of neighbors is about 30.
For a smaller radius the neighbor list is updated more often,
while for a larger radius future collisions are found among a
larger set of possible interactions.

A random insertion procedure gives an {\em equilibrated} configuration for the ideal gas, since the radial distribution function $g(r)$ is flat. But the dynamics retains a long memory and the structural relaxation slows down exponentially with density. For this reason it is more efficient to perform the averaging by choosing statically independent starting configurations and running them on different cores. For most of our simulations we use 512, 1728, and 4096 particles with $\rho$ varying from 1 to 30. All the simulations up to $\rho = 20$ and $N=4096$ are run for at least $10^9$ collision times or until the smallest nonzero wave vector $q=2\pi{V}^{-1/3}$ component of the dynamic structure factor $S(q,t)/S(q,0)$ has decayed to $1/e$, whichever is smaller. The runs with $\rho > 20$ and $N > 4096$ are not sufficiently long to satisfy the second condition, so they are only used to determine quantities measured on shorter time or length scales. The averaging procedure employed still guarantees the validity of these results. Finite-size effects are found to be negligible for all static and two-point quantities in the density regime under study, but four-point correlations exhibit sizeable size dependence. This will be discussed further in Sec.~\ref{sec:chi_4}.

\section{Results}
\subsection{Collisions and Diffusion\label{sec:res:bulk}}
\begin{figure}
\centerline{\includegraphics[width=\columnwidth, clip=true]{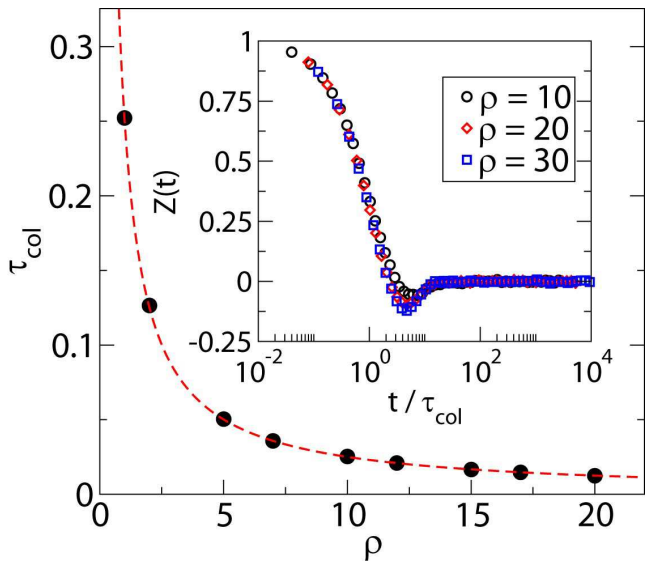}}
\caption{(Color online) Average collision time $\tau_{\mathrm{col}}$ from simulation (points) compared to the the kinetic theory prediction (dahsed line). Inset : Collapse of the velocity autocorrelation function $Z(t)$ after rescaling time by $\tau_{\mathrm{col}}$.}
\label{fig:collision}
\end{figure}

By construction the static properties of the system are those of an ideal gas. Moreover, for the density range considered the short-time dynamics agrees within errors with a mean-field kinetic theory. Fig.~\ref{fig:collision} shows in fact that the average time between rod collisions is indistinguishable from the analytical prediction $\tau_{\mathrm{col}} = 4/(9 \rho \pi^{1/2})$ (see Appendix). After only a few collisions the particle velocities become nearly completely uncorrelated, as gathered from the velocity autocorrelation function $Z(t) = \langle\mathbf{v}_j(t) \cdot \mathbf{v}_j(0) \rangle  / \langle|v|^2\rangle $ (Fig.~\ref{fig:collision} inset).  The nearly perfect collapse of $Z(t)$ after rescaling time by $\tau_{\mathrm{col}}$ in Fig.~\ref{fig:collision} shows this process to be rather general. The small negative dip of $Z(t)$ that follows at high density is the caging signature and corresponds to the bouncing back of a particle after colliding with a neighbor.  As far as structure and short-time dynamics are concerned the system thus behaves rather ideally.

\begin{figure}
\centerline{\includegraphics[width=\columnwidth,
clip=true]{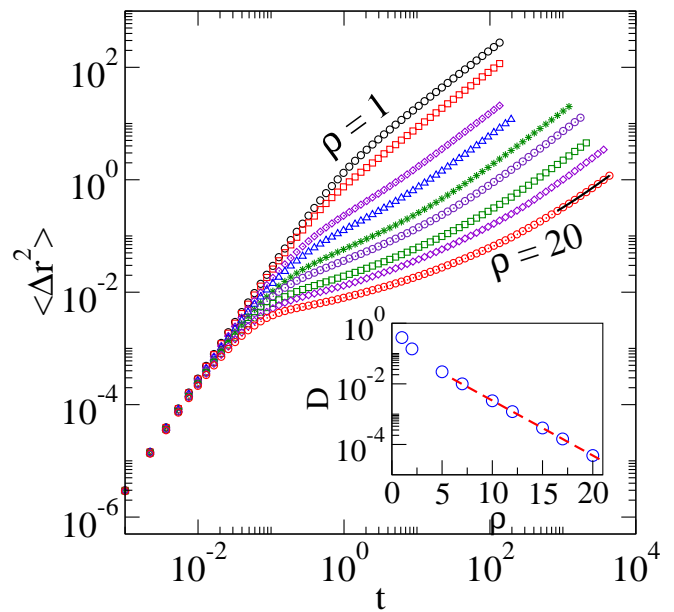}} \caption{(Color online) Time evolution of the MSD for $\rho$= 1, 2, 5, 7, 10, 12, 15, 17, and 20, from left to right. Superimposed to the long time part of $\rho = 20$ is a linear fit whose slope is used to calculate the diffusion coefficient. The error is smaller than the symbol size. Inset : The diffusion coefficient decreases with density exponentially. The dashed line is a fit $D\sim\exp[-\Delta V^*\rho]$ with $\Delta V^*=0.42$ for $\rho>5$.}
\label{fig:msd}
\end{figure}

On longer timescales the physics is quite different. Fig.~\ref{fig:msd} shows that the mean-square displacement (MSD) [$\langle\Delta r^2(t)\rangle \equiv \langle|\mathbf{r}_j(t) - \mathbf{r}_j(0)|^2\rangle$] between the initial ballistic regime [$\langle\Delta r^2(t)\rangle \sim t^2$] and the diffusive regime [$\langle\Delta r^2(t)\rangle \sim 6 D t$], where $D$ is the diffusion coefficient, develops a plateau for increasing densities as in supercooled fluids. But contrary to structural liquids there is no upper limit to packing, so the transition away from the ballistic regime takes place at ever shrinking length and time scales with increasing density. Instead of converging at a single length scale set by the repulsive core, as is the case in structural glass formers, the crossover plateau thus keep lowering with the slowdown. With the end of the plateau region, the system enters the diffusive regime on a timescale that grows exponentially with density. This suggests that the rate-limiting step for diffusion is the creation of  ``free volume'' around a particle, such that the topological constraints inhibiting its motion are relieved. For an ideal gas the probability to open up a volume $\Delta V^*$ by a spontaneous fluctuation is $\sim\exp(-\rho \Delta V^*)$. The exponential density dependence of $D$ thus suggests that a cavity with volume $\Delta V^* \simeq 0.42 \sigma^3$ is needed to enable diffusion. This behavior is very different from the algebraic density dependence observed for the rotational diffusion in systems of tethered, rotating needles~\cite{renner:95}. It is also unlike that of structural athermal systems such as hard spheres, where a power law is observed at modest undercooling~\cite{dawson:03}.  Exponential slowing down is more akin to what is obtained in strong glass formers.

\subsection{Self-intermediate scattering function}
\label{sec:fsqt}
\begin{figure*}
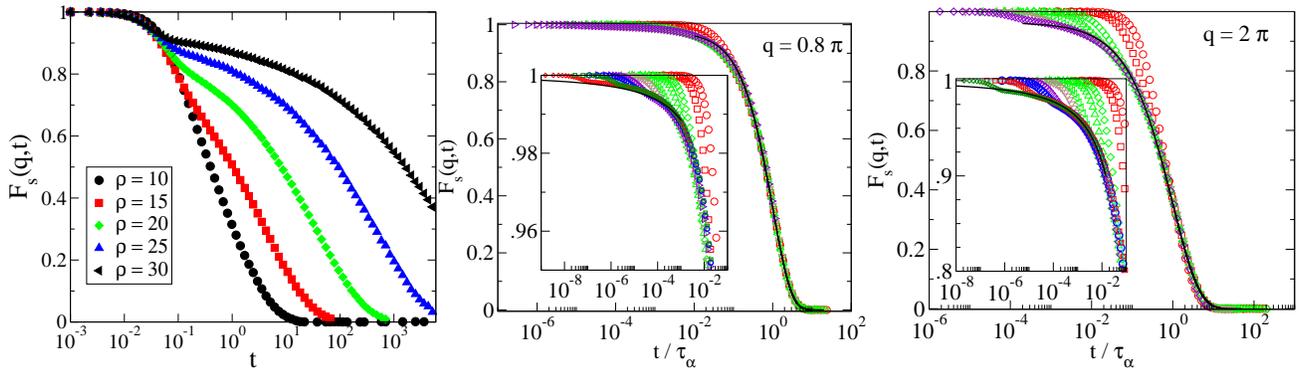

\centerline{\includegraphics[width=\switchsize\columnwidth,
clip=true]{Fs1_2.eps}\includegraphics[width=\switchsize\columnwidth,
clip=true]{das_etal_07.eps}\includegraphics[width=\switchsize\columnwidth,
clip=true]{das_etal_08.eps}}\caption{(Color online) $F_s(q,t)$ decay at (a) the microscopic wave vector $q_{\mathrm{cage}}$ as well as its $\tau_{\alpha}$ collapse at (b) $q=0.8 \pi$ and (c) $q=2\pi$ for $\rho$ = 5, 7, 10, 12, 15, 17, and 20. The solid line is a stretched exponential fit to $A\exp[-(t/\tau_\alpha)^\beta]$ for $F_s(q,t)$ between 0.025 and 0.975 with (b) $A$=0.975, $\beta$=0.917 and (c) $A$=0.963, $\beta$=0.662. Insets: Short time decay of $F_s(q,t)$ with additional $\rho$=22, 25, and 30. High-density ($\rho>20$) estimates for $\tau_{\alpha}$ are obtained by forcing the early $\alpha$ decay onto the master curve from the low-density data. The solid line is the free-particle decay form with $2.4\tau_{\mathrm{col}}$, as described in the text.}
\label{fig:fsqt_tau}
\end{figure*}

\begin{figure*}
\centerline{\includegraphics[width=\switchsize\columnwidth,
clip=true]{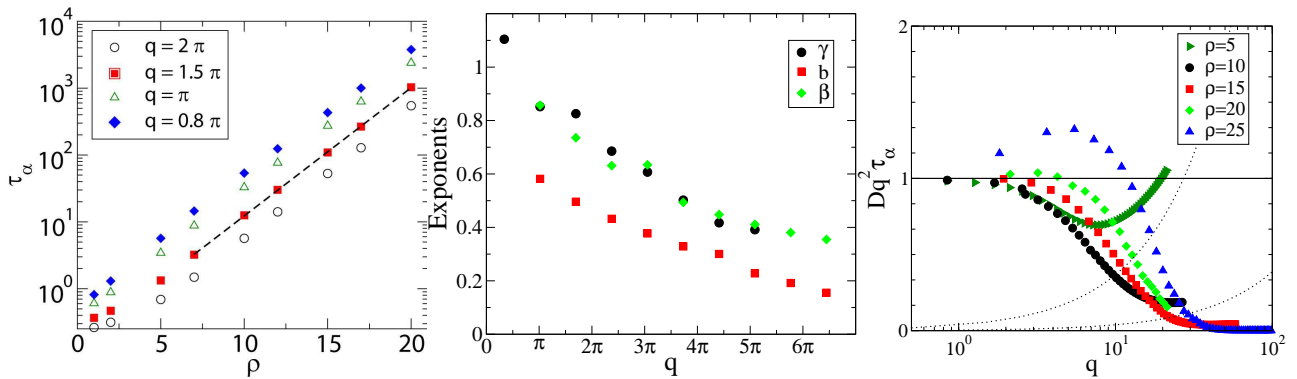}\includegraphics[width=\switchsize\columnwidth,
clip=true]{mu_q_2.eps}\includegraphics[width=\switchsize\columnwidth,
clip=true]{SE_break.eps}} \caption{(Color online) (a) Structural relaxation time $\tau_{\alpha}$ extracted from $F_s(q,t)$ for various wave vectors. The dashed line is an exponential fit to $q = \pi$ with exponent $0.43$.  (b) Exponents extracted from $F_s(q,t)$ and $\chi^{\rho}_4(q,t)$ at $\rho=20$, as described in the text of Sec.~\ref{sec:res:bulk} and Sec.~\ref{sec:chi_4}. (c) Rescaling of $\tau_{\alpha}(q)$ by the diffusive limit $Dq^2$ to evaluate the transport coefficient decoupling. The solid line emphasizes the Fickian limit $Dq^{2}\tau_{\alpha}(q)=1$, while the dashed lines show the small wavevector limit $\sqrt{2}Dq$ for the two lowest densities.}
\label{fig:tau_alpha}
\end{figure*}

The decay of density fluctuation on different length scales is best studied by the incoherent self-intermediate scattering function $F_s(q,t) \equiv \langle\frac{1}{N}\sum_j \exp\{i\mathbf{q}\cdot [\mathbf{r}_j(t) - \mathbf{r}_j(0)] \} \rangle $, where $q \equiv |\mathbf{q}|$ as reported in Fig.~\ref{fig:fsqt_tau}. In standard glass formers this correlation function bears the signature of two different dynamical regimes in the microscopic relaxation. On times of the order of $\tau_{\mathrm{col}}$, ballistic motion gives way to the $\beta$ plateau associated with caging; on longer times scales, $\alpha$ structural rearrangements allow a particle to escape the cage formed by its neighbors. The typical timescale $\tau_{\alpha}(q)$ over which this last process takes place is defined as the time when $F_s(q,t)$ has decayed to $1/e$. Here, $\tau_{\alpha}$ increases exponentially with density (Fig.~\ref{fig:tau_alpha}a). This supports the assumption that an infinite cross density is necessary to obtain complete dynamical arrest. The length scale at which caging and structural relaxation are best separated is the caging diameter. In structural glass formers it also corresponds to the first peak of the structure factor, but since the crosses do not exhibit any static structure,we approximate it instead by the average spacing between particles $q_{\mathrm{cage}}\equiv 2\pi\rho^{1/3}$. The growing separation between the two timescales with density can be observed in Fig.~\ref{fig:fsqt_tau}a. However, in spite of there being a difference of over three orders of magnitude between $\tau_{\mathrm{col}}$ and $\tau_{\alpha}$ at $\rho=20$, the apparition of a plateau is still incomplete. Higher densities are necessary to observe a better delineated structure.

An analysis of the structural relaxation process shows that when time is rescaled by $\tau_{\alpha}(q)$ for a fixed $q$, $F_s(q,t)$ collapse onto a single master curve with increasing accuracy as the system gets denser (Fig.~\ref{fig:fsqt_tau}b-c). This time-density scaling of the long-time decay is highly non-trivial. It has been argued that the collapse of the $\alpha$-relaxation curves is one of the outstanding characteristics of the structural glass transition that is reproduced by MCT \cite{gotze:91,gotze:91b}. Standard MCT being here inapplicable for lack of static correlations the phenomenon is clearly more generic. Long-time relaxation in glassy systems are often described by a stretched exponential Kohlrausch-Williams-Watts (KWW) form $F_s(q,t)\sim e^{-(t/\tau_{\alpha})^\beta}$, where the stretching exponent $\beta$ is not to be confounded with the $\beta$-relaxation regime. The long length scale limit is properly captured at low wave vectors, as seen in Fig.~\ref{fig:tau_alpha}b. The stretching exponent $\beta$ then approaches unity. This corresponds to an exponential decay of $F_s(q,t)$ and is consistent with the diffusive dynamics of simple fluids.  At microscopic length scales the KWW fit is also rather successful, as shown in Fig.~\ref{fig:fsqt_tau}b-c, though no single parametrization of the functional form is suitable for the entire decay range~\cite{ketel:05}. In particular, the fitting form does not capture the long-time tail, which falls off faster than expected from a fit to the body of the decay. For the bulk of the decay however, a singular behavior is observed. In other model glass formers such as silica~\cite{berthier:07} and binary Lennard-Jones (LJ)~\cite{kob:1995}, $\beta\gtrsim 0.75$ for wavevectors around $q_{\rm cage}$. For the crosses the decay is further stretched with $\beta\approx0.5$. This suggests that the structural relaxation arises from a broader characteristic-time distribution of relaxation processes.

MCT further predicts that the end of the $\beta$ plateau bends down following a von~Schweidler form ~\cite{gotze:91,gotze:91b}
\begin{equation}
F_s(q,t) \simeq f_c(q) - B h(q) (t/\tau_{\alpha})^b,
\label{eq:vonSchweidler}
\end{equation}
where $f_c(q)$ is the plateau height and both $B$ and $h(q)$ are independent of time. Equation~\ref{eq:vonSchweidler} approaches a stretched exponential form in the large-$q$ limit~\cite{fuchs:94}. For densities considered here this form is not obviously appropriate, since no convincing plateau has yet developed. This leaves $f_c(q)$ as a free fitting parameter to extract the exponent $b$ from the decay shoulder at $\rho=20$, as reported in Fig.~\ref{fig:tau_alpha}b. Though coarse, this treatment will be useful when we return to this issue in Sec.~\ref{sec:chi_4}.

A feature not part of the canonical glass analysis is the short-time collapse of $F_s(q,t)$, as presented in Ref.~\cite{ketel:05} and depicted in the insets of Fig.~\ref{fig:fsqt_tau}b-c. At short times the particles' ballistic movement leads to an initial Gaussian decay of $F_s(q,t)$. This regime ends when the ``free'' crosses collide with the ``cage'' formed by their neighbors at time $\tau_{\mathrm{col}}$ on average. Using
\begin{equation}
F^{\mathrm{free}}_s(q,\tau_{\mathrm{col}})=\exp{(-k_B T q^2\tau_{\mathrm{col}}^2/2m)}\label{eq:fsfree}
\end{equation}
and the scaling of $\tau_{\alpha}$ with density, one can parameterically plot where the change of regime from ballistic to collisional should take place for various densities. Since this does not correspond directly to a particular feature of $F_s(q,t)$, let's consider a larger value than $\tau_{\mathrm{col}}$ to describe the observed change in regime. Mobile particles have more free space around them (see Sec.~\ref{sec:dyn_het}) and contribute longer to the free decay of $F_s(q,t)$, so this is not unreasonable. Equation~\ref{eq:fsfree} with $2.4\tau_{\mathrm{col}}$ indeed captures the regime change at early times, as seen the insets of Fig.~\ref{fig:fsqt_tau}b-c. This time parameter is close to the first zero of $Z(t)$, another metric for the onset of caging. This explanation is rather system specific, so this collapse is not expected to be observed in other glass-forming systems.

\subsection{Dynamical heterogeneity \label{sec:dyn_het}}

\begin{figure*}
\centerline{\includegraphics[width=\switchsize\columnwidth,clip=true]{tailsnorm20.eps}
\includegraphics[width=\switchsize\columnwidth,clip=true]{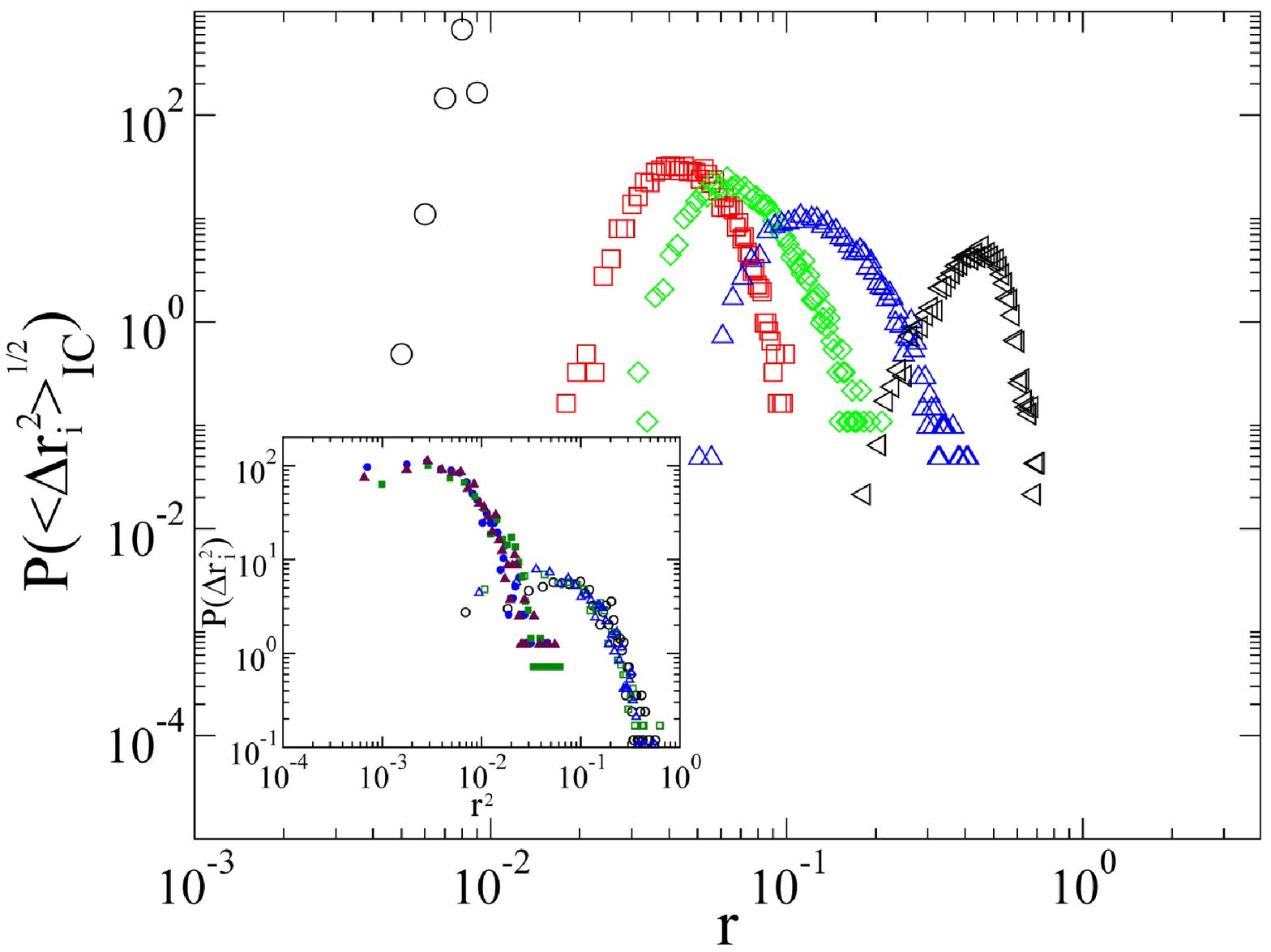} \includegraphics[width=\switchsize\columnwidth,clip=true]{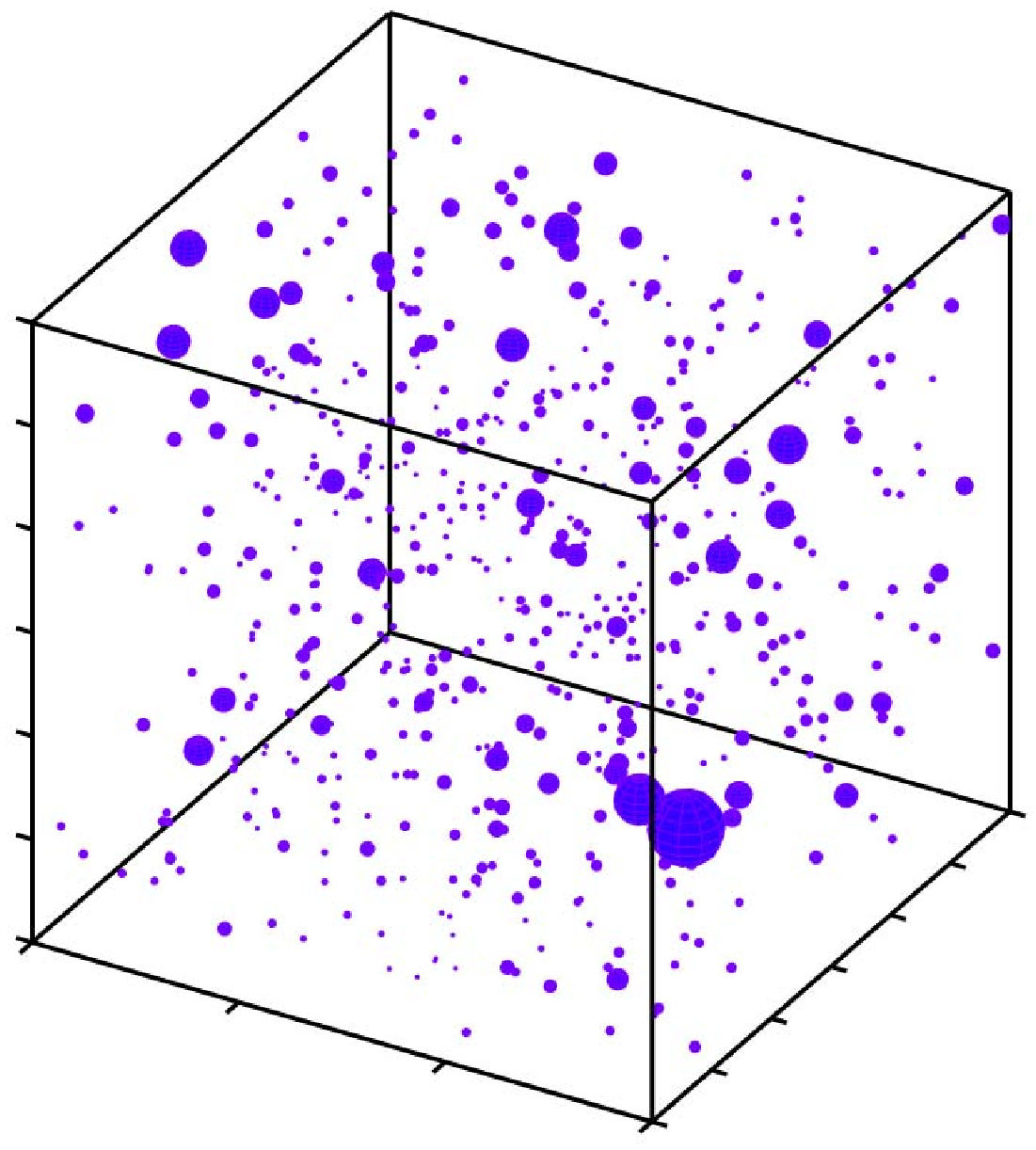}} \caption{(Color online) For a system at $\rho=20$. (a) Displacement probability distribution with superimposed Gaussian fits $P(r(t)) = 4\pi r^2(t) \left(2\pi \langle r^2(t)\rangle/3\right)^{-3/2} \exp(- 3 r^2(t)/2 \langle r^2(t)\rangle)$ for $t= 5.4 \times 10^{-3}$, $0.059$, $0.39$, $15$, $580$, and $3.6 \times 10^3$, from left to right. Arrows point to the excess probability for particles with large displacements.  Inset: One-dimensional component of the displacement probability at $t=32\approx\tau_{\alpha 2}$ with a Gaussian and an exponential fit small and large amplitudes respectively. (b) Propensity probability distribution $P(\langle \Delta r^2_i\rangle_{\mathrm{IC}}^{1/2})$ for the same first five times. Inset: distribution of displacements for the $0.07\%$ particles with the largest (open symbols) and smallest (closed symbols) propensities at $t=33$. (c) Propensities at $t=33$ shown as spheres centered around the initial particle positions. The spheres have a radius 2.5 times the magnitude of the individual particle propensities. The box has a side $3\sigma$.
}\label{fig:disp_heter}
\end{figure*}

\begin{figure*}[htbp]
\centerline{
\includegraphics[width=0.63\columnwidth,clip=true]{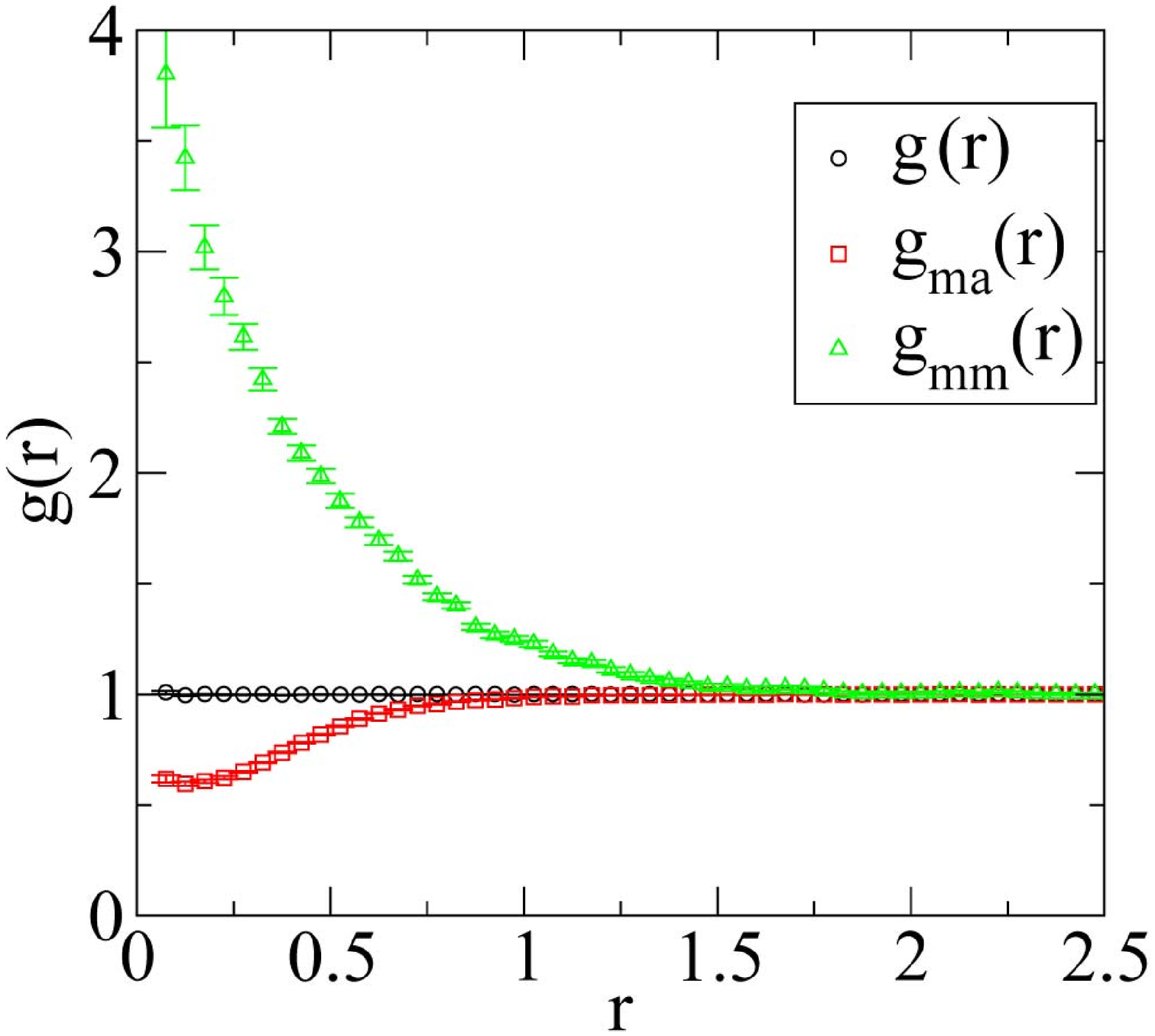} \includegraphics[width=\switchsize\columnwidth,clip=true]{das_etal_17.eps} \includegraphics[width=\switchsize\columnwidth,clip=true]{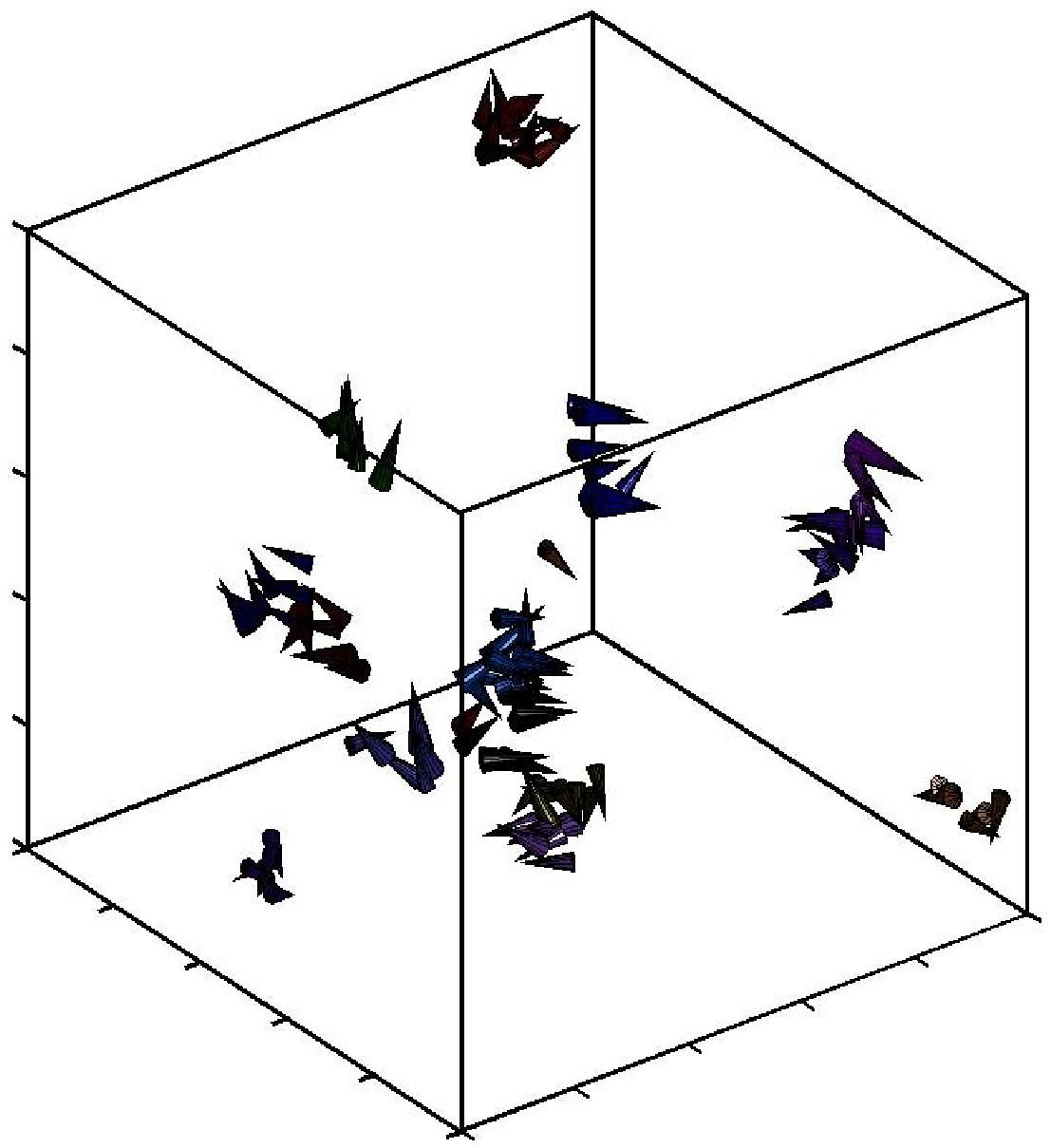}} \caption{(Color online) Mobility analysis for a system of $20,000$ crosses at $\rho=20$. (a) The radial distribution function $g(r)$ is compared with the conditional distributions $g_{ma}(r)$ and $g_{mm}(r)$. (b) Correlation of the mobile particle displacement directions, as described in the text.  (c) Displacements of clustered mobile particles over $\tau_{\alpha_2}$. The cone's base is at a mobile particle's initial position and the cone's height is twice its squared displacement over $\tau_{\alpha 2}$. Different shadings code for independent clusters. The box has side $10\sigma$.} \label{fig:mob20k_gr}
\end{figure*}

Various transport properties correspond to different moments of the distribution of microscopic times, so their decoupling at a particular wavevector is associated with the growth of dynamical heterogeneity on the corresponding length scale~\cite{berthier:2004,berthier:2005}. We first probe this effect using the wavevector dependence of $\tau_\alpha(q)$ rescaled by $D q^2$ and then looking for the onset of decoupling $Dq^2\tau_\alpha(q)>1$. As seen in Fig.~\ref{fig:tau_alpha}c, for small wavevectors the Fickian limit $D q^2\tau_\alpha(q)=1$ is recovered, while at very high wavevectors the Gaussian decay of $F_s(q,t)$ leads to a trivial $\sqrt{2} Dq$ growth. The transition from one regime to the other takes place over microscopic sizes $q\gtrsim q_{\mathrm{cage}}$. In denser systems decoupling is more pronounced and takes place at increasing length scales. For the highest densities the onset of decoupling suggests that particles have a coherent dynamics over distances as large as $4-5\sigma$. This is similar to what is observed in binary LJ under similarly sluggish relaxation~\cite{whitelam:2004,berthier:2004}, but here the number of particles involved is here an order of magnitude larger. We will come back to this issue in Sec.~\ref{sec:chi_4}, but note for now that since this size scale corresponds to the box dimension at these densities, it sets a computational upper bound to the range of densities reasonably accessible through simulations.

A number of simulation~\cite{kob:97,donati:98,yamamoto:98,donati:99,puertas:04} and experimental~\cite{marcus:99, weeks:02} studies of glass-forming systems also show a close relationship between the non-Gaussian behavior of particle displacements and dynamical heterogeneity. At high densities the dynamics of the crosses is indeed heterogeneous: only a small fraction of all particles is responsible for a significant fraction of the total MSD between the ballistic and the diffusive regimes, where the MSD plateaus. The probability distribution of particle displacements in Fig.~\ref{fig:disp_heter}a shows a tail at high displacements for intermediate times, while at short and long times the distribution tends towards a Gaussian. Deviations can be quantified using higher-order cumulants, the simplest of which is the fourth-order $\alpha_2(t) \equiv \frac{3 \langle r^4(t)\rangle }{5 \langle r^2(t)\rangle^2} - 1$. It vanishes when a distribution is truly Gaussian, but for the crosses and for structural glass formers it peaks more prominently and at longer times with increasing density. At $\rho = 20$ and time $\tau_{\alpha_2}$, when the non-Gaussian parameter $\alpha_2(t)$ reaches its maximum value, only 5\% of the particles are responsible for nearly $30\%$ of the MSD~\cite{ketel:05}.

To look further in the microscopic features of this phenomenon, mobile and slow particles have to be identified. The distinction between the two types is not a sharp one and depends on the time interval under consideration. For both short and long time intervals all particles have a similar MSD and the labels lose their meaning altogether. Kob {\em et al}. define a critical value of the displacement at a given time beyond which the self part of the van~Hove function deviates significantly from the corresponding Gaussian approximation~\cite{kob:97}. Particles that have a displacement larger than this critical value are termed mobile. This distinction between the two regimes can be observed near $\tau_{\alpha_2}$ in the inset of  Fig.~\ref{fig:disp_heter}a, where the short-range Gaussian and long-range exponential separation suggested in Ref.~\cite{chaudhuri:2007} captures the data reasonably well. From a different point of view, Shell~{\em et al}. showed that the joint probability distribution of initial velocity component and displacement along the same direction can be fitted by the sum of two Gaussian functions at intermediate times~\cite{debenedetti:05}. These authors suggest that the relative weights of the Gaussian components can be used to estimate the fraction of particles that are respectively mobile and immobile on that timescale. In this study we find that different measures of heterogeneity yield essentially the same results near $\tau_{\alpha_2}$. For this reason we use a simpler prescription: the $5\%$ of particles with maximum displacement at $\tau_{\alpha_2}$ are termed mobile~\cite{donati:99}.

In order to understand the physical origin of dynamical heterogeneities, it is important to gain insight in the factors that make a particular particle mobile. One possibility is that the distance over which a particle moves is sensitive to the initial velocity of that particle, but $Z(t)$ decays so rapidly that this could hardly be the whole story. Alternatively, the future mobility of a particle can be related to the detailed geometry of its initial local environment~\cite{cooper:04,cooper:04b,cooper:04c,cooper:2007,appignanesi06}. To distinguish between the two we consider an ``ensemble'' of trajectories that initiate from the same starting configuration, but with different initial velocities. Simulations of such an ``iso-configurational ensemble'' (IC) allow to determine whether the propensity for high mobility is related to the initial velocity or to the initial structure. If the former holds different particles are mobile from one trajectory to another, while if the latter holds the identity of mobile particles is correlated over different trajectories~\cite{cooper:04,cooper:04b,cooper:04c,cooper:2007}. For this we define the particle's propensity to diffuse at time $t$ as the IC average of the square of its displacement $\langle \Delta r_i^2 (t) \rangle_{\text IC} \equiv \langle |\mathbf{r}_i(t) - \mathbf{r}_i(0)|^2 \rangle_{\text IC}$. At both short and long times the distribution is expected to tend towards a delta function, because all starting positions are equivalent. Before a collision takes place all heterogeneities are kinetic, while for $t\gg\tau_{\alpha_2}$ all possible environments are sampled. If there is a structural contribution to dynamical heterogeneity it should thus appear at intermediate times. We run replicates of identical starting configurations at $\rho = 20$ to look at the distribution of propensities.  Figure~\ref{fig:disp_heter}b shows that removing the spread due to kinetic effects indeed gives a thinner propensity distribution than the full displacement distribution of Fig.~\ref{fig:disp_heter}a. But the relative width of the displacement distribution still grows until $t\sim\tau_{\alpha_2}$ and decreases afterwards. There is thus a structural component to dynamical heterogeneity in the cross system. However, though some particles have a propensity much higher than others, no feature of the distribution allows for a separation between propensity regimes, contrary to structural glass formers~\cite{cooper:04}. To see if there is nonetheless speciation, we look at the displacement distribution for the extremes of propensity. Yet in spite of having an average propensity an order of magnitude apart, their displacement distributions at $t\sim\tau_{\alpha_2}$ still overlap (Fig.~\ref{fig:disp_heter}b inset). Thus only a probabilistic propensity categorization is possible at the particle level.  But as for structural glass formers, it could still indicate that certain regions of space are structurally more mobile than others~\cite{berthier:07b}. We consider this option in Fig.~\ref{fig:disp_heter}c, where the spatial distribution of particle propensities at $\tau_{\alpha_2}$ is depicted as spheres centered around the initial particle position. It is hard to properly assess the regions of higher mobility directly from this representation. Though there appears to be some mobile ``domains'', where the most highly mobile particles can be found, these are not very large and for the rest the mobile particles appear more or less uniformly distributed over the system. This is significantly different from the large regions of similar propensity that are observed in structural glass formers~\cite{cooper:04,cooper:04b,cooper:04c,cooper:2007}. Either dynamically heterogeneous regions are here much smaller or propensity is an insufficient microscopic observable to capture their essence in crosses.

We nonetheless examine quantitatively possible spatial correlations among mobile particles with eight instances of a system of $20,000$ crosses at $\rho = 20$. The radial distribution function distinguishing the mobile particles from the rest is compared to the featureless system-wide $g(r)$ in Fig.~\ref{fig:mob20k_gr}a. The conditional probability of finding {\em any} particle at a distance $r$ given that a mobile particle is located at the origin $g_{ma}(r)$ shows a depression near $r=0$. This indicates that mobile particles tend to be found in local low-density regions, as suggested by the relaxation mechanism presented in Sec.~\ref{sec:res:bulk}. The radial distribution function of mobile particles alone $g_{mm}(r)$ indicates that they are also spatially correlated. It appears from this that mobile particles do organize in clusters over extended volumes. A different measure of correlations in the mobile particle distribution considers the displacement directions of mobile particles. For this we define a correlation function
\begin{equation}
O_m (r) \equiv \frac{\left\langle \Delta \mathbf{r}_m (0) \cdot
\Delta \mathbf{r}_m (r) \right\rangle }{\left\langle  |\Delta
\mathbf{r}_m|^2 \right\rangle },
\end{equation}
where $\Delta \mathbf{r}_m (0)$ is the displacement over the time interval $\tau_{\alpha_2}$ of a mobile particle considered to be at the origin and $\Delta \mathbf{r}_m (r)$ is the displacement of mobile particles in a spherical shell of radius $r$. Without correlations among the displacement direction of mobile particles $O_m (r)$ would be zero, while a non-zero value indicates some degree of assistance between mobile particles. Fig.~\ref{fig:mob20k_gr}b shows a positive $O_m$ at small $r$, so mobile particles' movements are only correlated when they are sufficiently close together to be ``entangled''. The negative dip that follows might be due to poor statistics, but this cannot be resolved here.

In structural glass formers mobile particles are sometimes found in clusters with a ramified morphology~\cite{donati:98}. The analysis done so far leaves open the possibility of chain-like movements for the cross model, which incites us to look directly at the spatial distribution of mobile particle clusters. Here, two mobile particles belong to a same ``cluster'' if their separation is less than $\sigma/2$ in all directions at both initial and final times. This threshold is similar to the decay length scale of $g_{mm}(r)$. It ensures that members of a cluster share collision history over the entire time interval during which displacement is considered. Most mobile particles do not belong to such a cluster and only 10\% of them belong to clusters of size six or more; the largest cluster identified contains 14 particles. Figure~\ref{fig:mob20k_gr}c shows clusters of six or more mobile particles as cones with a base centered around the particles' initial position and oriented along their displacement. We find no indication of non-compact or linear chains of mobile particles contrary to what was observed in simulations of the binary LJ glass former~\cite{donati:98,donati:99}. This allows to conclude that high-mobility clusters do indeed exist and that they are not only small, but also compact. But at such high density, though the system has undergone a significant dynamical slowdown, collectively relaxing regions remain of limited spatial extent.

\subsection{Dynamical susceptibility \label{sec:chi_4}}
\begin{figure*}
\centerline{\includegraphics[width=\switchsize\columnwidth]{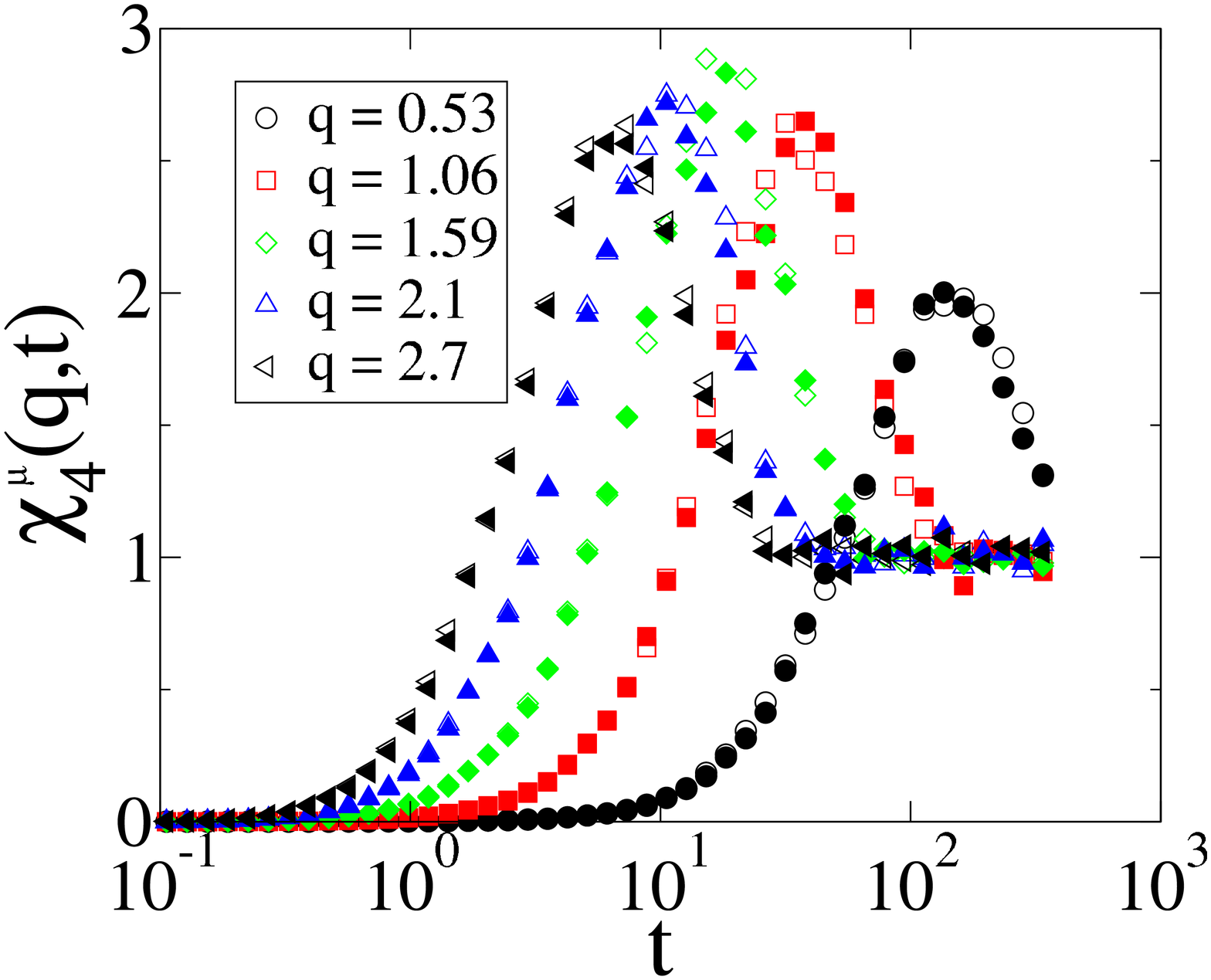}
\includegraphics[width=\switchsize\columnwidth]{chimax_size_inset.eps}
\includegraphics[width=\switchsize\columnwidth]{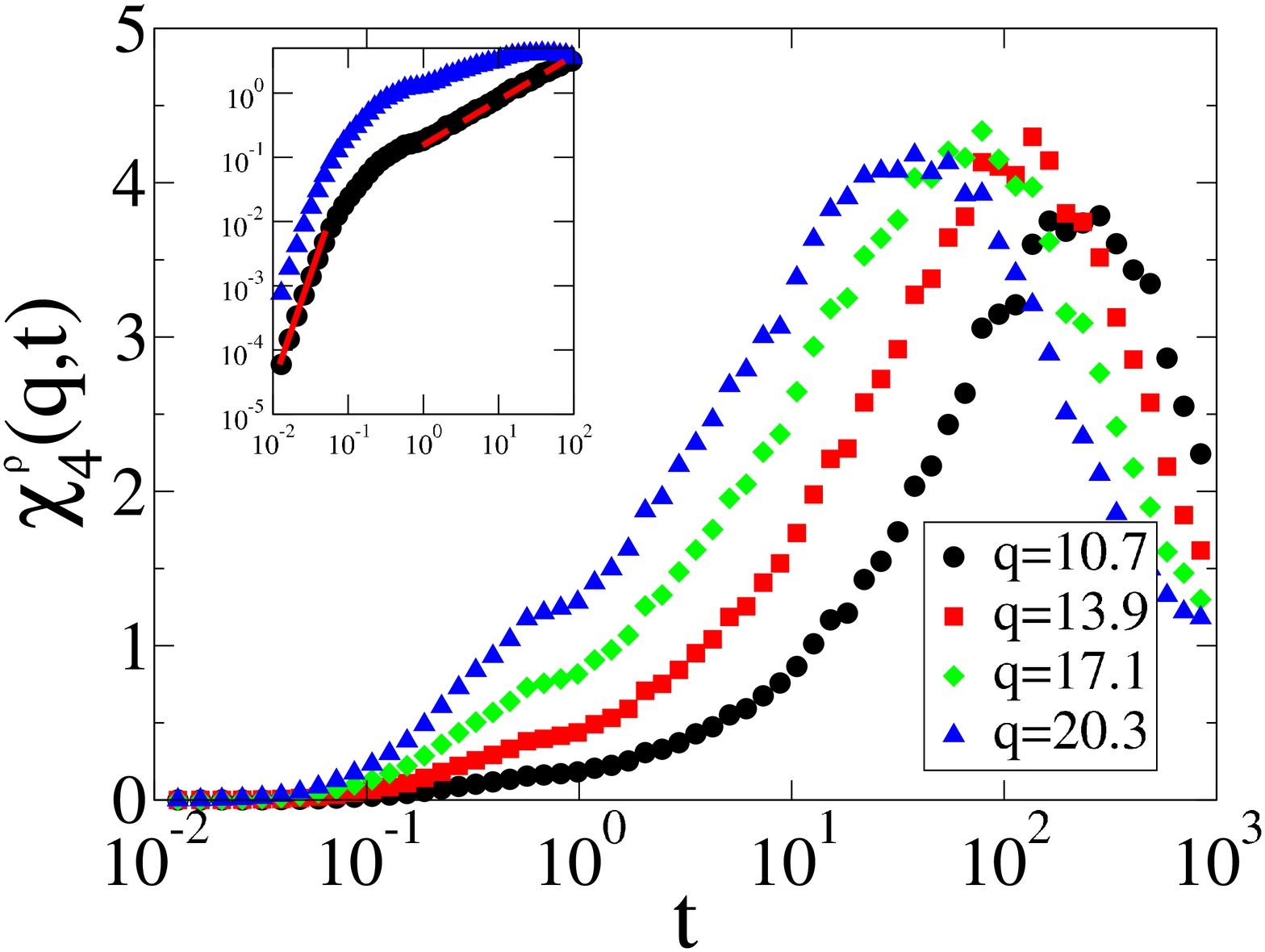}}
\caption{(Color online) (a) Determination of  $\chi_4^{\mu}(q,t)$ with $\langle N\rangle=8192$ at $\rho=5$ by two different approaches: direct simulation (full symbols) and through Eq.~\ref{eq:chi4mu} (empty symbols). (b) System-size dependence of $\chi_4^{\rho *}$ at $\rho=5$. Inset: $\chi_4^{*}$ for $\langle N\rangle=8192$ at constant $\rho$ (squares) and constant $\mu$ (circles). (c) Dynamic susceptibility $\chi^{\rho}_4(q,t)$ for $\rho = 20$. Inset: The solid line $t^4$ follows the ballistic behavior at $t<\tau_{\mathrm{col}}$ and the dashed one the intermediate power-law regime.}\label{fig:chi4}
\end{figure*}

A particularly useful quantity to discriminate between different models of dynamical arrest and to provide further information about the relaxation mechanism is the four-point density correlator
\cite{dasgupta:91,donati:99:chi4,glotzer:00,toninelli:05,berthier:05,chandler:06,szamel:06}
\begin{eqnarray}
G_4(\mathbf{r}, t) &=& \langle \Delta  \rho(0,0) \Delta  \rho(0,t) \Delta  \rho(\mathbf{r}, 0) \Delta  \rho(\mathbf{r}, t) \rangle  \nonumber \\
&&- \langle \Delta \rho(0,0)\Delta  \rho(0,t) \rangle  \langle \Delta \rho(\mathbf{r}, 0)\Delta  \rho(\mathbf{r}, t) \rangle ,
\end{eqnarray}
where $\Delta \rho(\mathbf{r}, t)$ denotes a density
fluctuation at position $\mathbf{r}$ and time $t$. $G_4$ probes
the spatial correlation in the decay of density fluctuations at
different times. The volume integral of $G_4(\mathbf{r}, t)$ is its associated susceptibility $\chi_4(t)$, which is also a measure of the variance of the correlation function $\langle \Delta \rho(0,0)\Delta \rho(0,t) \rangle$. Numerical simulations show that the information contained in this reduced dynamic susceptibility is very similar to the full four-point density correlator~\cite{glotzer:00}. In practice it is convenient to compute a phase-space correlator in terms of the self-intermediate scattering function
\begin{equation}
f_s(\mathbf{q}, t) \equiv \frac{1}{N} \sum_{j=1}^N e^{i\mathbf{q} \cdot \left[\mathbf{r}_j(t)-\mathbf{r}_j(0)\right]}.
\end{equation}
From this definition we recognize that $F_s(q,t)=\langle f_s(\mathbf{q},t)\rangle$. In athermal systems, the corresponding dynamic susceptibility is then
\begin{equation}\label{eq:chi4}
\chi_4^{\rho}(q,t) = N \left[ \left\langle  f_s(\mathbf{q},t)^2 \right\rangle_{\rho} - \left\langle  f_s(\mathbf{q}, t) \right\rangle^2_{\rho} \right]
\end{equation}
at constant density. We use the $\rho$ label, because unlike for the two or the full four-point correlators the susceptibility depends on the choice of simulation ensemble~\cite{berthier:05,berthier:2007,chandler:06}. The ``true'' susceptibility is obtained by keeping the chemical potential $\mu$ fixed instead. This can be done directly or using the derivative of the two-point function
\begin{equation}\label{eq:chi4s}
\chi_4^{\mu}(q,t)=\chi_4^{\rho}(q,t)+\rho k_BT\kappa_T\left(\frac{\partial F_s(q,t)}{\partial \ln{\rho}}\right)^2_T,
\end{equation}
where $\kappa_T$ is the isothermal compressibility and $\mu$ refers to the constant chemical potential. For crosses with $k_BT=1$ it reduces to
\begin{equation}\label{eq:chi4mu}
\chi_4^{\mu}(q,t)=\chi_4^{\rho}(q,t)+\left(\frac{\partial F_s(q,t)}{\partial \ln{\rho}}\right)^2_{T}.
\end{equation}
This result is tested for $\langle N\rangle=8192$ at $\rho=5$ in Fig.~\ref{fig:chi4}a. We use the scheme described in the Appendix of Ref.~\cite{chandler:06} on the one hand and numerical differentiation of the two-point function added to $\chi_4^{\rho}(q,t)$ on the other. The two approaches agree with each other within numerical uncertainty. We can estimate the difference between the two ensembles from the inset of Fig.~\ref{fig:chi4}b. At small $q$ around $\tau_{\alpha}$ the two-point correction is similar in magnitude to $\chi_4^{\rho}$, but the density fluctuation term becomes negligible for wavevectors larger than $q_{\mathrm{cage}}$. This is consistent with the results from facilitated models~\cite{chandler:06}.

The main panel of Fig.~\ref{fig:chi4}b shows a prime feature of the dynamic susceptibility: its peak height $\chi_4^{\rho *}(q)$. It corresponds to the maximum in dynamical heterogeneity on a given length scale and thus takes place on times of the order of $\tau_{\alpha}(q)$. Surprisingly we find $\chi_4^{\rho*}$ to have appreciable system-size dependence even for a density as low as $\rho=5$.  At higher densities these effects are also pronounced, but their study becomes rapidly computationally intractable. Transport coefficients analysis in Sec.~\ref{sec:fsqt} did suggest that a dynamical length scale might be as large as the box size for $\rho\gtrsim 20$. But even for a system 16 times larger than the typical size considered so far and at much lower density, $\chi_4^{\rho}(q,t)$ has not yet converged to its bulk value. Considering $\chi_4^{\mu*}$ does not change this observation. Also, not only does $\chi_4^{\rho*}$ keeps increasing with system size, but it keeps shifting to smaller wavevectors. There thus exists a dynamical length scale in this system that is much larger than the system size, even at densities where caging barely interferes with the diffusive regime. Moreover this takes place as the peak height, which scales with the dynamical heterogeneity volume, remains modest.  Such large scale dynamical heterogeneity could result from low-amplitude, long-range fluctuations of the two-point correlation, since their integration over a large volume would give them a prominent contribution. This could then blur the details of local dynamical heterogeneity normally associated with a dynamical slowdown. Whatever its origin, this effect prevents us to quantify completely the wavevector dependence of $\chi_4^{\rho*}(q,t)$ or the scaling of its peak height, as was done in Refs.~\cite{chandler:06,charbonneau:07}. A comment remains nonetheless in order. The broad distribution of wavevectors over which the peak of $\chi_4^{\rho*}$ develops indicates that relaxation processes leading to structural relaxation take place over a range of length scales. Because no single microscopic scale dominates, the mean-field cage opening picture for diffusion might be more caricatural than in structural glass formers. Many different microscopic mechanisms are probably at play, as the small value of the stretching exponent had already suggested in Sec.~\ref{sec:fsqt}.

For microscopic $q$ finite-size effects are less important, so we will only consider these smaller length scales to test theoretical predictions on the other properties of the dynamical susceptibility.
The full time and wave-vector dependence of $\chi_4^{\rho}(q,t)$ shows a rich structure~\cite{toninelli:05,chandler:06,charbonneau:07}. At short times the motion is ballistic $\chi_4^{\rho}(q,t)\sim t^4$, it exhibits a maximum at $t^*(q)$ close to the structural relaxation time $\tau_{\alpha}(q)$, and at long times goes to unity. Between the ballistic regime and the peak the function is often fitted to a power-law $\chi_4^{\rho}(q,t) \sim t^{\gamma(q)}$, since theoretical predictions for $\gamma(q)$ differ depending on the dynamical relaxation mechanism involved. If short-lived events are responsible for the loss of correlations $\gamma = 1$, for independently diffusing defects $\gamma= 2$, while MCT predicts it to be the same as the exponent $b$ from the von~Schweidler form of Eq.~\ref{eq:vonSchweidler}. This last scenario is observed in the binary LJ glass former~\cite{toninelli:05}, but numerical results for kinetically constrained models are consistent with the assumption of diffusive point-like defects with an anomalous diffusion exponent~\cite{chandler:06}.

Well-separated power-law regimes and the peak of $\chi_4^{\rho}(q,t)$ can be seen in Fig.~\ref{fig:chi4}c. The exponent $\gamma$, obtained for wave vectors where the power-law growth lasts at least one time decade at $\rho=20$, depends strongly on $q$ (Fig.~\ref{fig:tau_alpha}b). To check the MCT prediction we compare $\gamma$ to exponent $b$ extracted from the fit to Eq.~\ref{eq:vonSchweidler}. The two exponents are significantly different from each other for all wavevectors. However, since the von Schweidler functional form does not satisfyingly describe the late $\beta$ regime even at the highest density considered this is not a conclusive assessment. Instead, because of the improperly-defined plateau $\gamma$ probably corresponds to exponent $\beta$ of the stretched-exponential decay, as field-theoretic arguments suggest~\cite{berthier:2007b}. Figure~\ref{fig:fsqt_tau}b presents a remarkable agreement between $\gamma$ and $\beta$, which support this interpretation. A clear separation between the von Schweidler and the KWW regimes develops only at densities higher than what is accessible through simulations, so it cannot be excluded that an additional power-law regime corresponding to the von Schweidler regime then be observed.

\section{Concluding remarks}
\label{sec:conclu} We have considered a system of particles formed by fixing three orthogonal line segments rigidly at their midpoints. Absence of excluded volume implies an absence of static correlations, so all the static and thermodynamic properties are that of an ideal gas. However, the non-crossing condition for the line segments gives rise to highly nontrivial dynamics and exhibits ``glassy'' features as the number density is increased. A volume needs to open up for a particle to diffuse away from its neighbor cage, and this activated dynamics makes the model a ``strong'' glass former. In spite of the inapplicability of standard MCT for this system we observe properties that are traditionally considered to be success of MCT, such as the rescaling of the stretched exponential relaxation in $F_s(q,t)$. It remains unclear why such predictions should hold here and if some of them break down at densities beyond what is computationally reasonable. Note also that a model with a similarly trivial static, but fragile glass-forming behavior, would also be of great interest to test the assumptions that underlie the categorization.

With increasing density particle displacements acquire strong non-Gaussian features on the structural relaxation timescale. During this time a small fraction of the particles show a much larger MSD than rest. We find these ``mobile'' particles to be associated with local low density regions and to cluster. However, the mobile clusters tend to be small and highly localized. Yet both the transport coefficient decoupling and the system-size dependence of the dynamical susceptibility indicate that a sizeable dynamical length scale is present in the system. In light of the mobility study and the magnitude of the dynamical susceptibility this comes as a surprise, because these are usually taken as indirect probes of the dynamical heterogeneity volume. The task to reconcile the large dynamical length scale with the small size of the mobile regions might require identifying a different microscopic metric for dynamical heterogeneity. Alternately, this behavior shows features that are reminiscent of elastic relaxation of a solid after a local volume change. Though this effect has not been observed in other glass-forming systems so far, it might have been obscured by a stronger local dynamical heterogeneity. In any case, a better understanding of this phenomenon would benefit the study of all glass-forming systems.

\begin{acknowledgments}
The work of the FOM Institute is part of the research program of FOM and is made possible by financial support from the Netherlands Organization for Scientific Research (NWO) and by EU contract No. MRTN-CT-2003-504712. Computer time at the Dutch center for high-performance computing SARA is gratefully acknowledged.  C.D.~acknowledges financial support from EPSRC and Soft matter composites (SOFTCOMP), while P.C.~acknowledges MIF1-CT-2006-040871 (EU) funding. We further gratefully acknowledge helpful discussions at different stages of this work with H.~C.\ Andersen, C.~A.\ Angell, J.-L.\ Barrat, L.~Berthier, M.~E.\ Cates, D.\ Chandler, M.\ Fuchs, W.~van~Ketel, H.\ L\"owen, D.~R.\ Reichman, L.\ van Rooyen, R.\ Schilling, F.\ Sciortino, M.\ Sperl, and T.\ Voigtmann.
\end{acknowledgments}

\appendix*

\section{Collision Frequency}
\label{sec:appendix} We take two needles of length $\sigma$. In
an interval of time $\Delta t$, the number of collisions for
these two needles is
\begin{equation}
\Gamma_{nn}=2 \rho v_\perp^{rel}\Delta t
\left|\sin{\theta}\right|\sigma^2,
\end{equation}
where $v_\perp^{rel}$ is the relative perpendicular velocity
and $\theta$ is the angle between the two line segments. The
factor of two appears because two lozenges of size
$\sigma^2\sin\theta$ are formed. The perpendicular
relative velocity averages to
\begin{equation}
\langle v_\perp^{rel}\rangle =
\left(\frac{k_BT}{8\pi m_r}\right)^{1/2}=\frac{1}{2\sqrt{\pi}},
\end{equation}
where $m_r$ is the reduced mass and the last equality follows
from using reduced units. Using
$\left\langle\left|\sin{\theta}\right|\right\rangle=\pi/4$
we get
\begin{equation}
\Gamma_{nn}=\rho\frac{\sqrt{\pi}}{4}.
\end{equation}
Since each cross is made up of three needles, an additional
factor of $9$ has to be included to get the cross collision
frequency that is used in the text
\begin{equation}
\Gamma_{cc}=9\times \Gamma_{nn}=\frac{9 \rho\sqrt{\pi}}{4}.
\end{equation}

\bibliography{glass}
\end{document}